\documentstyle[aas2pp4]{article}

\received{10 October 1997}
\accepted{TBD}
\slugcomment{To appear in ApJ number NN}

\lefthead{L.\ Loinard \& R.J.\ Allen}
\righthead{Cold Massive Molecular Clouds in M31}

\newcommand{\HII}{\mbox{H\,{\sc ii}}}
\newcommand{\HI}{\mbox{H\,{\sc i}}}

\newcommand{\Htwo}{H$_{2}$}

\newcommand{\twCO}{$^{12}$CO}
\newcommand{\thCO}{$^{13}$CO}

\newcommand{\degrees}{$^{\circ}$}
\newcommand{\pcmcub}{\mbox{${\rm cm^{-3}}$}}
\newcommand{\kmps}{\mbox{${\rm km\;s^{-1}}$}}

\newcommand{\Msun}{\mbox {M$_{\scriptsize \odot}$}}

\newcommand{\Tkin}{T$_{\rm kin}$}
\newcommand{\Tb}{T$_{\rm b}$}
\newcommand{\Tex}{T$_{\rm ex}$}
\newcommand{\Xunit}{cm$^{-2}$ (K km s$^{-1}$)$^{-1}$}
\newcommand{\etamb}{$\eta_{\rm mb}$}
\newcommand{\al}{$\alpha$(1950.0)}
\newcommand{\be}{$\delta$(1950.0)}
\newcommand{\rg}{R$_{\rm gal}$}

\newcommand{\dechms}[4]{$#1^{\rm h}#2^{\rm m}#3\mbox{$^{\rm s}\mskip-7.6mu.\,$}#4$}
\newcommand{\intdms}[3]{$#1^{\circ}#2'#3''$}

\newcommand{\pwr}[2]{\mbox{$#1 \times 10^{#2}$}}

\newcommand{\da}[1]{$\Delta\alpha =$#1$''$}
\newcommand{\dd}[1]{$\Delta\delta =$#1$''$}
\newcommand{\lsim}{\mbox{$\mathrel{\vcenter{\hbox{\ooalign{\raise3pt\hbox{$<$}\crcr \lower3pt\hbox{$\sim$}}}}}$}}
\newcommand{\gsim}{\mbox{$\mathrel{\vcenter{\hbox{\ooalign{\raise3pt\hbox{$>$}\crcr \lower3pt\hbox{$\sim$}}}}}$}}

\begin{document}

% TITLE PAGE

\title{Cold Massive Molecular Clouds in the Inner Disk of M31}

\author{Laurent Loinard\altaffilmark{1} and R. J. Allen}
\affil{Space Telescope Science Institute, 3700 San Martin Drive,
   Baltimore, MD 21218, USA}

\altaffiltext{1}{Observatoire de Grenoble, Laboratoire d'Astrophysique, 
  Universit\'e Joseph Fourier, B.P. 53X, F-38041 Grenoble Cedex, France}

% ABSTRACT

\begin{abstract}

We present new interferometric \twCO(1-0) and single-dish \twCO(3-2)
observations of the central parts of D478, a large ($> 200$ pc) dark
dust cloud located in a quiescent region of the inner disk of M31 where
single-dish \twCO(1-0) and \twCO(2-1) observations were previously
obtained. Only a small fraction ($ < 15\%$) of the \twCO(1-0) flux
previously detected in this region with the single-dish telescope is
recorded by the interferometer. Most of the \twCO(1-0) emission must
therefore have the appearance of a smooth surface with very little
structure on scales smaller than $\approx 25''$ (85 pc).  Together with
the earlier \twCO(1-0) and \twCO(2-1) single-dish results the new
\twCO(3-2) data are in good agreement with LTE predictions for
optically thick lines at \Tex\ = \Tkin\ = 3.5K. These results rule out
the conventional model for these clouds consisting of warm clumps with
a low filling factor (as would be the case if they resembled Galactic
GMCs) and confirm that large, massive, cold molecular clouds exist in
the inner disk of M31 with kinetic temperatures close to that of the
cosmic microwave background. Such extremely low temperatures are likely
to be a consequence of the low heating rate in these particular regions
of M31, where very little massive star formation is occurring at
present. 

From the \twCO\ line profile widths we estimate the Virial mass surface
density of D478 to be 80 -- 177 \Msun\ pc$^{-2}$. This is a factor 7 -- 16
times larger than the value obtained by multiplying the \twCO\ profile
integrals with the conventional ``X-factor''.

\end{abstract}


\section{Introduction}\label{intro}

Unlike the annular region of prominent star formation at 8 kpc $<$
\rg\ $<$ 11 kpc, the inner disk of M31 (\rg\ $<$ 7 kpc) contains very
little of the classical tracers of massive star formation such as
\HII\ regions and young clusters (e.g.\ Baade \& Arp 1964). It is also
nearly devoid of emission from the usual tracers of the cool
interstellar medium such as \HI\ (e.g.\ Brinks \& Shane 1984) and CO
(e.g.\ Dame et al.\ 1993). This situation is remarkable because prominent
dust ``arms'' are visible throughout the inner disk of M31 (Baade
1963), and hundreds of dust clouds have been catalogued in these
regions by Hodge (1980a, 1980b). Should we conclude that there is no
interstellar gas associated with these dust clouds? Could all the gas
be frozen out on grain surfaces? Or is the gas present but just in a
physical state which we cannot easily detect?

Using the IRAM 30-m radio telescope, Allen \& Lequeux (1993) and
Loinard, Allen \& Lequeux (1996) detected faint but broad \twCO(1-0)
and \twCO(2-1) profiles associated with several dust clouds in the
inner disk of M31. Although these clouds obey the same size -- line
width relationship as Galactic Giant Molecular Clouds (GMCs), thereby
suggesting similar masses, their \twCO(1-0) luminosities are typically
only 1/10 that of Galactic GMCs with the same line widths, and their
\twCO(2-1)/\twCO(1-0) line intensity ratios are unusually low. Two of
these clouds, designated D268 and D478 in the catalog by Hodge (1980a),
were subsequently observed in \thCO\ with the 30-m telescope (Allen \&
Lequeux 1994); the \twCO/\thCO\ ratio ($\approx 10$) confirmed that, in
spite of their faintness, the \twCO\ lines are optically thick.
Equilibrium chemistry and radiative transfer calculations (Allen et
al.\ 1995) further showed that the observed CO brightnesses are
consistent with optically-thick lines emerging from a very
low-temperature extensive medium. Such low temperatures can be
understood as resulting from a lack of obvious heat sources (UV photons
and cosmic ray primaries) in the inner disk of M31, a likely
consequence of the very low rate of massive star formation there.

An alternative model for the observed CO emission in the inner disk of
M31 follows from detailed studies of Galactic GMCs. In that picture the
CO emission emanates from a collection of small, warm clouds,
unresolved by the 30-m single-dish observations. The faintness of the
CO emission results from the small area filling factor within the beam
of the 30-m telescope.  The large line widths would then have to be
ascribed to random motions of these clouds driven by some process which
is not related to the mass derived from the virial theorem. Because
warm, thermalized clouds should have \twCO(2-1)/\twCO(1-0) line ratios
of $\approx 1.0$, the observed values of \lsim\ 0.5 would have to be a
result of subthermal excitation conditions in a medium of low volume
density.  Unlike the case for cold, extended gas, this interpretation
leads to the conclusion that only small amounts of molecular gas are
present in the inner disk of M31.

We have carried out two new complementary observing programs in order
to test these contrasting views. First, using the
Millimeter Wave Array at the Caltech Owens Valley
Radio Observatory\footnote{The OVRO is supported in part by NSF Grant
96-13717 and the K.T. and E. L. Norris Foundation.} (OVRO), we have
obtained an aperture synthesis image of a $\approx 1'$ field within
D478 at a resolution of $\approx 5''$ in the \twCO(1-0)\ line.  If the
CO flux detected by the 30-m observations is coming from spatially
compact components, they ought to be easily visible in the synthesis
image with peak brightnesses of several Kelvins. If, on the other hand,
the emission is mostly smooth, little or none of it will be recorded in
the OVRO synthesis image. Second, we have obtained \twCO(3-2) spectra at
the center of D478 and at a reference position in the bright
star-forming ring of M31 with the James Clerk Maxwell
Telescope\footnote{The JCMT is operated by the Royal Observatories on
behalf of the Particle Physics and Astronomy Research Council of the
United Kingdom, the Netherlands Organisation for Scientific Research,
and the National Research Council of Canada.} (JCMT). Owing to its
higher critical density, the 3-2 transition should enable us to
determine whether subthermality is the cause of the low \mbox{2-1/1-0}
ratios.

\section{Observations}\label{obs}

Radial velocities in this paper will refer to the Local Standard of
Rest (LSR); for comparison with other observations made in the
Heliocentric system (HEL) we note that, at the position of M31, these
systems are related by V$_{\rm HEL}$ = V$_{\rm LSR} - 4.3$ \kmps.
Positions are measured as offsets $\Delta\alpha = 15 \times (\alpha -
\alpha_0) \times \cos(\delta)$ and $\Delta\delta = \delta - \delta_0$
from the nominal center of M31 at \al\ = \dechms{00}{40}{00}{3}, \be\ =
\intdms{40}{59}{43}.

\subsection{The OVRO observations}

The interferometer observations were taken with the 6-element
Millimeter Wave Array at OVRO between 1996 December 8 and 1997 January
28. The spectrometer was a digital autocorrelator providing 32 MHz of
available bandwidth at a spectral resolution of 1 MHz; at 115 GHz, this
yields a velocity coverage of 83 \kmps\ and a velocity resolution of
2.6 \kmps.  A single field of view (HPBW $\approx 65''$) was observed
centered on the position of D478 as defined by Allen \& Lequeux (1993)
at offset position \da{+245}\ and \dd{+473}.
The data were calibrated at Caltech using the
software system developed by Scoville et al.\ (1993) for the Millimeter
Wave Array. The instrumental passband was derived from integrations on an
artificial noise source obtained at the beginning of each track.  Phase
and amplitude gains were calibrated using observations of 0133+476
obtained every half-hour throughout the observations, and the absolute
flux density scale (accurate to 20\%) was calibrated using observations
of Uranus and Neptune obtained at the beginning and/or end of each
track.  Corrections for atmospheric absorption were made continuously
using the standard chopper-wheel method, which is also used at the IRAM
30-m telescope. After calibration, the flux density we recorded for
0133+476 is 2.0 Jy at 115 GHz, in excellent agreement with the values
obtained for that source with the 30-m telescope at the end of 1996
(Ungerechts 1997, private communication).  This gives us confidence
that the flux density scales at OVRO and at the 30-m telscope are the
same to within about 10\%.

The OVRO data were imaged in {\it AIPS} using natural weighting of the
visibilities and some tapering, CLEANed to the $1\sigma$ level (50 mJy
beam$^{-1}$ $\approx$ 0.15K), and restored with an elliptical gaussian
beam of $6.8'' \times 4.6''$ FWHM (PA = -27\degrees). The OVRO
array does not record short-spacing visibility data; the shortest
baseline obtained during our observing run was 15 meters which,
allowing for forshortening during the observation, provided visibility
data down to $\approx 3490 \lambda$. A source smaller than $\approx\,
10''$ will therefore be essentially fully recorded in our OVRO
visibility measurements ($V > 0.90$), but a source larger than $50''$
will be virtually absent ($V < 0.10$)\footnote{More precisely, when
observed with an interferometer of baseline $u$ (measured in
$\lambda$), a 2D symmetric gaussian source of FWHM $\theta$ (in
arcseconds) will be recorded with a visibility amplitude of $V =
\exp{(-8.37 \times 10^{-11}u^2\theta^2)}$. At $\theta = 26''$, $V \approx
0.5$.}.  The restoration process will therefore produce channel maps
with an estimate of that part of the source brightness distribution
which has structure on angular scales \lsim\ $25''$.

\subsection{The JCMT observations}\label{smooth}

The CO(3-2) spectra were obtained with the JCMT on 1995 July 20.  At
345 GHz, the HPBW of the JCMT is $15''$, and the main beam efficiency
\etamb\ = 0.58. The data were obtained in position-switching mode, and
the backend was the {\it Dwingeloo Autocorrelation Spectrometer}
providing 760 MHz of available bandwidth at a spectral resolution of
756 kHz; at 345 GHz, this yields 650 \kmps\ of velocity coverage with a
resolution of $\approx 0.5$ \kmps. However, because the passband of the
frontend was only $\approx 700$ MHz, the useful range was limited to
about 600 \kmps. Since the observed profiles are about 15 -- 30
\kmps\ wide, the data were Hanning-smoothed to 2 \kmps\ to enhance the
signal-to-noise ratio. We observed both the center of D478, at
\da{+245}\ and \dd{+473}, and a reference position ``M31ref'' in the
bright star-forming ring at \da{+1359}, \dd{+679}. These two positions
were previously observed in the \twCO(1-0) and \twCO(2-1) lines with
the 30-m telescope (Allen \& Lequeux 1993, Loinard et al. 1996).

During the previous 30-m observations, we obtained data at positions in
a small 5-point cross with $12''$ spacing. The central spectrum was
observed twice, once at the beginning of the cycle (hereafter S1), and
once at the end (hereafter S6). In between, we obtained spectra at the
4 flanking fields (hereafter S2, S3, S4 and S5). The 6 spectra can be
combined to smooth the CO(2-1) to different resolutions, using the
equation S = [0.5(S1+S6)+$\alpha$(S2+S3+S4+S5)]/(4$\alpha+1$). With
$\alpha = 0.5$, the 30-m CO(2-1) spectra can be smoothed to the
resolution of the 30-m telescope at 115 GHz ($23''$, corresponding to
75 pc), and with $\alpha = 0.17$ to the $15''$ resolution of the JCMT
at 345 GHz.

\section{Results and Discussion}\label{resu}

\subsection{Structure of the CO(1-0) emission in the OVRO images}

All of the emission detected in the OVRO synthesis is confined to the 9
central image channels, as shown in Figure \ref{chanmaps}. Note that
the circles drawn on each channel map indicate the $23''$ beam of the
30-m telescope at 115 GHz, not the $65''$ OVRO field of view.  A double
source is clearly resolved in channels b -- d at the center of the map,
and an additional elongated source is seen in the north-west part of
the field in channels f -- i. Figure \ref{chanmaps}j shows contours of
the total CO(1-0) emission detected in the OVRO maps; in Figure
\ref{dust+co} these contours are drawn over an optical image  of this
part of M31 obtained from the Digitized Sky Survey\footnote{The optical
image is from a short-exposure V plate taken with the Palomar Schmidt
telescope and digitized at the Space Telescope Science Institute under
U.S. Government grant NAG W-2166. The Oschin Schmidt Telescope is
operated by the California Institute of Technology and Palomar
Observatory.}. The contours of the emission detected at OVRO are elongated
in the same direction as an underlying dust feature, but the CO contours
are significantly narrower than the dust lane.

\subsection{The fraction of the 30-m flux in the OVRO map}

After smoothing the OVRO channels maps to the $23''$ resolution of the
IRAM 30-m telescope at 115 GHz, we corrected these maps for
attenuation by the primary beam of the array elements, and extracted
the velocity profiles at the positions we had previously observed with
the 30-m telescope. The results are shown in Figure \ref{spectra}. The 
thin-lined histograms in this Figure show the 30-m data, and the thick
lines the OVRO data. Representative gaussian fits are also shown. The
spectrum labelled ($+245'';+473''$) corresponds to the circles drawn at
the centers of the channel maps in Figure \ref{chanmaps}; the spectrum
at ($+245'';+497''$) is located a full beamwidth to the north.

It is clear from Figure \ref{spectra} that {\it only a small fraction
of the 30-m flux has been recovered in the OVRO maps}. At the
central position ($+245'';+473''$) the 30-m flux is 66.5 Jy \kmps, but
the integral under the OVRO spectrum is only 9.9 Jy \kmps\ or 15\% of
the 30-m flux. At the northern position ($+245'';+497''$) the 30-m
flux is 68.5 Jy \kmps, whereas the OVRO flux is 5.1 Jy \kmps\ or only 
7.5\% of the 30-m flux. It is also clear that the OVRO profiles are
qualitatively different from those recorded by the 30-m.  The component
near -85 \kmps\ in Figure 3 apparently has $\sim 10 - 30\%$ of its
energy in spatial structures \lsim\ $25''$, whereas the component near
-75 \kmps\ must be very smooth on these angular scales.

\subsection{Excitation conditions from the 30-m and JCMT spectra}

We shall see momentarily that the CO lines emanating from D478 are
consistent with high optical depth.
For optically-thick lines, the observed brightness temperature
\Tb\ is related to the excitation temperature \Tex\ by: 
\begin{equation}
T_b = T_0 \biggr(\frac{1}{e^{T_0/T_{ex}}-1}-\frac{1}{e^{T_0/T_{bg}}-1}\biggr)
\label{lte}
\end{equation}
\noindent
where $T_{bg} = 2.73$K is the temperature of the cosmic microwave
background (Mather et al. 1994), and $T_0 = h\nu/k$, where
$\nu$ is the frequency of the observed transition. 

We have re-analyzed the 30-m \twCO(1-0) and the \twCO(2-1) spectra first
reported by Allen \& Lequeux (1993) using an improved method of
combining the individual observations, i.e.\ weighting by the inverse
square of the actual r.m.s.\ noise in each spectrum rather than by the
integration time.  The results (at $23''$ resolution) are shown in
Figure \ref{d478}a. The spectra differ slightly from the original
results published in Figure 2b of Allen \& Lequeux, although the
differences are minor. The peak on our revised \twCO(1-0) spectrum is
now $\sim 0.05$K lower, while the \twCO(2-1) spectrum peak is $\sim
0.07$K higher\footnote{Note also that Allen \& Lequeux used Heliocentric
velocities.}. In Figure \ref{d478}b we show the \twCO(2-1) data at $15''$
and add our new JCMT \twCO(3-2) result. The procedure used to obtain the
smoothed profiles is described in \S\ref{smooth}.

Since for Galactic GMCs the area filling factor over the observing beam
is generally $< 1$, it is usual to estimate the excitation temperature
from the ratios of two CO lines (assuming the filling factor is the
same for both lines). From Figure \ref{d478}a the \mbox{2-1/1-0} line
ratio for the -85 \kmps\ component at the center of D478 is 0.5, for
which Equation \ref{lte} yields an excitation temperature of \Tex\ =
3.5K. For this excitation temperature, Equation \ref{lte} predicts a
\twCO(1-0) brightness temperature of 0.6K, and a \twCO(2-1) brightness
temperature of 0.3K, which are exactly the peak temperatures observed.
We conclude that the filling factor within a $23''$ area is $\approx 1$,
consistent with the conclusion drawn earlier from the OVRO data that
most of the \twCO(1-0) emission from D478 emanates from a smooth
surface. The narrow ridge of emission in the OVRO map of Figure
\ref{dust+co} is indicative of a slightly warmer region; it appears to
be just resolved in the $5''$ OVRO map, with \twCO(1-0) peak brightness
temperatures of 1 -- 1.2K in a $5''$ beam, corresponding to an
excitation temperature of $\approx 4$K.

Figure \ref{d478}b shows the 30-m \twCO(2-1) profile (dashed line)
smoothed to the $15''$ resolution of the JCMT. We note that the
amplitude and general shape of this profile is nearly the same as that
of the $23''$ \twCO(2-1) profile in Figure \ref{d478}a, which is also
consistent with a filling factor $\approx 1$. For \Tex\ = 3.5K,
Equation \ref{lte} predicts that the \twCO(3-2) brightness temperature
should be 0.1K, in excellent agreement with the observed JCMT spectrum
(Figure \ref{d478}b). From this we conclude that the \twCO(3-2) line is
also thermalized, and that the excitation temperature in D478 is equal
to the kinetic temperature of the molecular gas.  The density of
the molecular gas is apparently of the order of several thousand
\pcmcub, the critical density for excitation of the CO(3-2) line.
{\it The combination of the OVRO \twCO(1-0) data with the 30-m and JCMT 
observations of the \twCO(1-0), \twCO(2-1) and \twCO(3-2) emission
lines is consistent with the conclusion that D478 is a large, smooth,
optically thick, thermalized molecular cloud at \Tkin\ = \Tex\
$\approx 3.5$K.}

We now compare the results just obtained in D478 with similar
observations of the reference position ``M31ref'' (\da{+1359},
\dd{+679}) in the bright star-forming ring of M31.
There, the excitation temperature deduced from the \mbox{2-1/1-0} ratio
(Figure \ref{m31ref}a) is 5K. But for \Tex\ = 5K, Equation \ref{lte}
predicts a \twCO(1-0) brightness temperature of about 1.9K; since the
observed \twCO(1-0) profile has a peak temperature of 0.75K, we expect
the CO filling factor inside the $23''$ beam of the 30-m telescope to
be $\sim 40\%$. Wilson \& Rudolph (1993) observed the field around this
position with the BIMA interferometer. They resolved the emission into
several small sources and recovered virtually all the single dish flux
density.  At \da{+1359}, \dd{+679}, they detected one molecular cloud
(M31-1, see their Figure 2), with size $14'' \times 17''$. The
corresponding filling factor in the $23''$ beam of the 30-m telescope
is then ($14 \times 17$)/($23 \times 23$) = 45\%, in excellent
agreement with what we have inferred from our observations.  However,
the JCMT data for this position in M31 can not be so readily
reconciled; the 3-2/2-1 line ratio (Figure \ref{m31ref}b) corresponds
(Equation \ref{lte}) to \Tex\ $\approx 9$K, much higher than the value
deduced from the \mbox{2-1/1-0} line ratio.  It is possible that for
this position, which is located in a region of massive star-formation,
embedded sources heat the cloud from the inside, thereby enhancing the
heating of the higher density regions. That no such additional heating
is present in D478 is consistent with the apparent lack of active star
formation in this region.

\subsection{Mass Surface Density}

It is clear that the conventional method of calculating the \Htwo\ mass
from the CO(1-0) brightness (as summarized e.g.\ by Allen 1996 and in
references given there) will fail for cold thermalized clouds like
D478, for the simple reason that the CO(1-0) brightness vanishes as the
cloud temperature descends towards that of the cosmic background at
2.73~K (cf.\ Equation \ref{lte}). This disappearance of the CO(1-0)
emission is, of course, quite independent of the total mass of
molecular gas present. We are left with the Virial theorem as the only
remaining method of estimating the underlying mass.  For this method to
work we need to have a tracer for the maximum extent of the turbulent
velocities present in the gas contained within the telescope beam. The
residual CO(1-0) emission profile remains useful for this purpose, even
for cold clouds, assuming that the parts of the clouds which are warm
enough to be detected have a range of velocities that is representative
of the turbulence in the entire area covered by the telescope beam. In
this picture, specific peaks in the total CO profile have no special
significance; only the total velocity extent of the whole profile
matters.

The virial mass $M_{vir}$ of a cloud of radius $R$ (in parsec),
integrated line width $\Delta v$ (FWHM in \kmps), and a constant
density distribution is (e.g.\ MacLaren et al.\ 1988) $M_{vir}/\Msun =
210 \times (\Delta v)^2 R$.  Since D478 fills the 30-m telescope beam
(FWHM = $23''$ or 75 pc), we can calculate the mass surface density at
e.g.\ the central position ($+245'';+473''$) in Figure \ref{spectra}.
At this position, the velocity profile width $\Delta v = 25$ \kmps.  In
the spirit of the model we are using, the fact that the profiles in
Figure \ref{spectra} appear to be separable into two components is
irrelevant. However, one correction which does need to be applied is to
account for any contribution to $\Delta v$ arising from a gradient of
the rotational velocity field of M31 over the $23''$ beam of the 30-m
telescope. This contribution can be estimated from the model velocity
field fitted to the 21-cm \HI\ synthesis of M31 by Brinks (1984, page
4-28); it is $\approx 4$ \kmps\ at the position of D478. This must be
linearly subtracted from the observed profile width, leaving $\approx
21$ \kmps. The mass surface density of D478 is therefore:

\begin{equation}
\Sigma \approx 210 \times \frac{(21)^2 \times \cos i}{(\pi \times 37.5)}
    \approx 177\; {\rm M}_{\scriptsize \odot} {\rm pc}^{-2}
\end{equation}

\noindent in the plane of M31, where we have taken the inclination of
M31 to be $i = 77$\degrees.

An alternative model is to consider the two peaks in the velocity
profile of D478 as two physically distinct ``clouds''.  We emphasize
that the structure of D478 which we have deduced from our own data,
namely, an extended cold cloud with a few localized warm regions on its
surface, {\em is not consistent with such an alternative model}.
Nevertheless, it is common to interpret velocity profiles in this way,
and we give the result here for completeness. In that case, each of the
two main components has $\Delta v \approx 10$ \kmps, and the mass
surface density at the center of D478 would then be:

\begin{equation}
\Sigma \approx 2 \times 210 \times \frac{(10)^2 \times \cos i}
{(\pi \times 37.5)}
    \approx 80\; {\rm M}_{\scriptsize \odot} {\rm pc}^{-2}.
\end{equation}

Finally, we note that the conventional method of determining masses
from CO profiles based on a CO-to-\Htwo\ ``conversion factor'' X =
\pwr{1.9}{20} \Xunit\ (e.g.\ Strong \& Mattox 1996) yields a mass
surface density of $\approx 11$ \Msun\ pc$^{-2}$ including a correction
factor of 1.36 for helium. This ``X-factor'' mass surface density is
about an order of magnitude less than that deduced from the virial
theorem. Magnani \& Onello (1995) also found large variations (a factor
10 or more) in the conversion factor both for translucent and for dark
clouds in the Galaxy. We conclude that this is an unreliable method of
determining molecular masses and variations in molecular mass surface
densities in galaxies.

\section{Concluding Remarks}

Our analysis of new \twCO(1-0) interferometric observations, and of
single-dish \twCO(1-0), \twCO(2-1) and new \twCO(3-2) observations,
confirms that the faint CO emission in the direction of D478 emanates
from an extended, smooth, massive structure at very low kinetic
temperature. These results are likely to apply to other dark clouds in
the inner disk of M31. In contrast, the CO emission from the
star-forming ring of M31 comes from smaller clouds with higher kinetic
temperatures, as is the case for Galactic GMCs.
 
These cold, massive, molecular clouds we have identified in the inner
disk of M31 appear to be quite different than Galactic GMCs. However,
the {\it observational} differences do not necessarily have to reflect
instrinsic structural differences, but could result from the same gas
finding itself in a different environment, as has been suggested by
Allen et al.\ (1995). Suppose the gas is an extensive, turbulent
medium with a wide range of densities. In the absence of strong
fluxes of UV photons (and cosmic rays) the low-density parts remain
molecular and cold, and this medium will appear faint, smooth and
extended in the CO lines, as is seen in D478. However, when subjected
to an intense flux of UV photons (as would be the case for gas located
near regions of massive star formation), the low density regions would
be dissociated and the remaining high-density regions will be heated,
so that the CO lines will appear to emanate from warm, high-density
clumps. This could explain the situation for Galactic GMCs, and also
for our reference position ``M31ref'' in the star-forming ring of M31.

The dust clouds in the inner disk of M31 constitute an excellent
``laboratory'' for the study of the large-scale physics of molecular
gas.  The relative scarcity of massive star-forming regions allows us
to observe molecular clouds in somewhat simpler situations. D478 for
instance, with its smooth appearance and apparent lack of embedded
sources, is apparently close to the idealized case of an infinite
plane-parallel cloud illuminated on one side by a flux of UV photons.
Studies of a larger sample of such clouds with varying local conditions
(nearby B stars, etc.) would be helpful in elucidating the interplay
between the ISM and star formation, and may also provide a more
complete view of the total molecular content of this galaxy.

\acknowledgements
We are very grateful to Dr.\ Nick Scoville for his encouragement and
advice on the observing program, and for discussions on the results. We
thank the OVRO time allocation committee for their award of observing
time on the Millimeter Wave Array, the OVRO staff for performing the
observations, and Dr.\ K.\ Sakamoto for his help with the data
reduction at Caltech. We are also grateful to Dr.\ R.\ Tilanus who
performed the \twCO(3-2) observations at the JCMT, to Dr.\ J.\ Lequeux
for his careful reading of the manuscript, and to our colleagues at the
Space Telescope Science Institute for stimulating discussions. The
Space Telescope Science Institute is operated by AURA for NASA under
contract NAS 5-26555. We especially thank the Director of the Institute
for supporting this research through an allocation from his
Discretionary Research Fund.

\clearpage

\onecolumn

\clearpage

\begin{figure}
\caption{\twCO(1-0) emission recorded by the OVRO Millimeter Wave Array
in the field of D478.  The center of the field is at offset \da{+245},
\dd{+473}\ from the nominal center of M31 at \al\ =
\dechms{00}{40}{00}{3}, \be\ = \intdms{40}{59}{43}. The circle in each
panel indicates the $23''$ beam (FWHM) of the 30-m telescope at 115
GHz.  The central (LSR) velocity of each channel is indicated in each
panel.  The maps have not been corrected for primary beam attenuation.
The synthesized beam is $6.8'' \times 4.6''$ FWHM (PA = -27\degrees)
and is shown in the lower left corner of panel (j).  Panels (a) to
(i):  The first contour and the contour interval are 100 mJy
beam$^{-1}$ = 0.3K ($\approx 2\sigma$).  Panel (j):  Total emission
detected by the OVRO array, obtained by integrating the channels (a) to
(i). The first contour and the contour level are 0.78 Jy beam$^{-1}$ \kmps.
The 5 crosses in this panel correspond to the positions of the 5
spectra in Figure \ref{spectra}.} \label{chanmaps}
\end{figure}

\begin{figure}
\caption{OVRO total \twCO(1-0) contours overlaid on a short exposure V image
from the Digitized Sky Survey. The first contour and contour interval
are 0.78 Jy beam$^{-1}$ \kmps. The circle indicates the size of
the OVRO field of view ($65''$ FWHM).}
\label{dust+co}
\end{figure}

\begin{figure}
\caption{\twCO(1-0) spectra at $23''$ resolution for 5 positions in the
region of D478. The panel labelled (+245'';+473'') is the central position
in Figure \ref{chanmaps}. The thin line histograms are the 30-m
single-dish profiles, and the thick lines are the smoothed OVRO
data. The y-axis is in Kelvins and is the same for all panels; the
X axis is LSR radial velocity.} \label{spectra}
\end{figure}

\begin{figure}
\caption{Panel (a): $23''$ resolution 30-m \twCO(1-0) (full line) and
\twCO(2-1) (dashed line) spectra at the center of D478 (\da{+245},
\dd{+473}). Panel (b): 30-m \twCO(2-1) (dashed line) spectrum
smoothed to $15''$ resolution of the JCMT,
and JCMT \twCO(3-2) (dotted line) spectrum at the same position.
The x-axis in both panels is LSR radial velocity.}
\label{d478}
\end{figure}

\begin{figure}
\caption{Same as Figure \ref{d478}, but for the reference position
  ``M31ref'' at \da{+1359}, \dd{+679}.}
\label{m31ref}
\end{figure}

\end{document}